\documentclass[journal,transmag]{IEEEtran}

\usepackage{cite}
\usepackage{graphicx}
\usepackage{amssymb,amsmath}
\usepackage{xcolor}
\usepackage{xspace}

\usepackage{amsmath}
\newcommand{\MM}[1]{\ensuremath{#1}\xspace}

\newcommand{\mv}[1]{\mathbf{#1}}

\newcommand{\frf}{\MM{f_\text{rf}}}

\begin{document}

\title{A semi-analytical model to simulate the spin-diode effect and accelerate its use in neuromorphic computing}


\author{
	\IEEEauthorblockN{Chloé Chopin\IEEEauthorrefmark{1}, Leandro Martins\IEEEauthorrefmark{2}, Luana Benetti\IEEEauthorrefmark{2}, Simon de Wergifosse\IEEEauthorrefmark{1}, Alex Jenkins\IEEEauthorrefmark{2},\\ Ricardo Ferreira\IEEEauthorrefmark{2}, and Flavio Abreu Araujo\IEEEauthorrefmark{1}}
	\IEEEauthorblockA{\IEEEauthorrefmark{1}Institute of Condensed Matter and Nanosciences, UCLouvain, Louvain-la-Neuve, Belgium, chloe.chopin@uclouvain.be}%
	\IEEEauthorblockA{\IEEEauthorrefmark{2}International Iberian Nanotechnology Laboratory, Braga, Portugal}%
}

\IEEEtitleabstractindextext{%
\begin{abstract}
The spin-diode effect is studied both experimentally and with our original semi-analytical method.
The latter is based on an improved version of the Thiele equation approach (TEA) that we combine to micromagnetic simulation data to accurately model the non-linear dynamics of spin-torque vortex oscillator (STVO). 
This original method, called data-driven Thiele equation approach (DD-TEA), absorbs the difference between the analytical model and micromagnetic simulations to provide a both ultra-fast and quantitative model.
The DD-TEA model predictions also agree very well with the experimental data.
The reversal of the spin-diode effect with the chirality of the vortex, the impact of the input current and the origin of a variation at half of the STVO frequency are presented as well as the ability of the model to reproduce the experimental behavior.
Finally, the spin-diode effect and its simulation using the DD-TEA model are discussed as a promising perspective in the framework of  neuromorphic computing.
\end{abstract}

}

\maketitle

\pagestyle{empty}
\thispagestyle{empty}

\IEEEpeerreviewmaketitle%
\section{Introduction}
\begin{figure}%
    \centering%
	\includegraphics[width=1.3in]{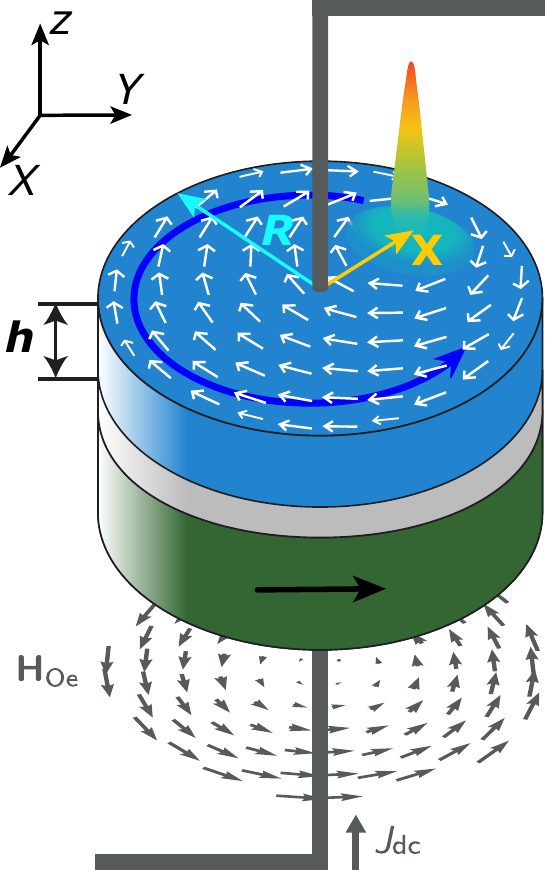}%
	\caption{A spin-torque vortex oscillator is represented with the polarizer in green (its magnetization is along the $y$-axis), the insulating layer in gray and the free layer in blue. The chirality of the vortex core is represented by white arrows. The AOF field generated by the input current is symbolized by gray arrows. Adapted from \cite{araujo2022data}}.%
	\label{fig:STVO}%
\end{figure}

%
\IEEEPARstart{S}{pin-torque} vortex-based oscillators (STVOs) are nanoscale devices with high potential for applications like radio-frequency (rf)-generation\cite{dussaux2010large}, rf detection\cite{markovic2020detection}, or neuromorphic computing\cite{grollier2020neuromorphic}.
They are based on the magnetic tunnel junction (MTJ) composed of two magnetic layers decoupled by an insulating spacer (see Fig.~\ref{fig:STVO}).
One of the magnetic layer is called the polarizer and has a fixed magnetization while the other magnetic layer is called the free layer as its magnetization can be freely controlled using a dc current for example. %
%
By carefully choosing the geometry of the free layer of the MTJ, a magnetic vortex can be nucleated as the nano-dot magnetization ground state\cite{guslienko2008magnetic}.
It has an in-plane curling magnetization except at the vortex core where it points out-of-plane.
Depending on the circulation direction of the magnetization which is either anti-clockwise or clockwise, a vortex has either a positive or negative chirality.
In addition, a vortex has a positive or negative polarity depending on the magnetization of the vortex core (either pointing up or down, respectively).
Its dynamics is impacted by the two topological parameters mentioned above, by the polarizer orientation and by the input current.
Furthermore, the latter generates an Ampère-Oersted field (AOF) which can not be neglected as it modifies the vortex core dynamics depending on its chirality \cite{abreu2022ampere, de2022quantitative}. %
%
A spin-diode effect\cite{tulapurkar2005spin} arises in a STVO when a rf current is injected into the device. 
The combination of the input current and the oscillation of magnetization of the MTJ due to the vortex core motion gives rise to an output oscillating voltage measured by tunnel magnetoresistance.
The dc component, which depends on the input current frequency, is then extracted (see Fig.~\ref{fig:SP_exp_vs_model}).
This phenomenon is suitable for applications like non-volatile memory\cite{martins2021non} or neuromorphic computing\cite{grollier2020neuromorphic}. %
%
The spin-diode effect can be easily measured experimentally and understood at some extend using micromagnetic simulations\cite{martins2021non}.
However, micromagnetic simulations are very time consuming and the results offer low resolution in terms of rf frequency in contrast to what is needed to explain the experimental features.
So, an original semi-analytical model called data-driven Thiele equation approach (DD-TEA)\cite{araujo2022data} is used to fill this gap. %
This DD-TEA model is a promising tool to study the spin-diode effect.
Here, its predictions are compared to experimental data and its potential use in neuromorphic computing is presented.%
%
%
%
\section{Methods}
Experimental data are obtained for a STVO with a radius of $R=500$~nm and a free layer with a thickness of $h=9$~nm.
The polarizer is along the $y$-axis (see Fig.~\ref{fig:STVO}). %
The DD-TEA model\cite{araujo2022data} is based on the TEA\cite{thiele1973steady} where the vortex is seen as a quasi-particle and the evolution of the vortex core position $\mv{X}$ is defined by %
\begin{equation}
    G(\mv{e}_z\times\dot{\mv{X}}) + D \dot {\mv{X}}= \frac{\partial(W^\text{ex} + W^\text{ms}+ W^\text{Oe}) }{\partial\mv{X}} + \mv{F}^\text{ST},
    \label{eq:thiele}
\end{equation}%
where $G$ and $D$ are the gyro-vector and the damping terms, $W^\text{ex},  W^\text{ms}, W^\text{Oe}$ are respectively the exchange, magnetostatic, and the AOF contribution to the potential energy, and $\mv{F}^\text{ST}$  is the spin-torque force.
Due to the assumptions used in TEA \cite{de2022quantitative}, only qualitative results are obtained\cite{abreu2022ampere}.
Thanks to a limited amount of micromagnetic simulations and our data-driven approach, the DD-TEA model allows to achieve both fast and quantitative results by absorbing the difference between the pure analytical model and micromagnetic simulations data\cite{araujo2022data}. %
%
The spin-diode effect is computed as follow
\begin{equation}
    \displaystyle
    \overline{\Delta V} \simeq \frac{\Delta R_{\text{STVO}}}{2}\frac{I_\text{ac}}{T_2 - T_1} \int^{T_2}_{T_1}\Delta m_y \sin(2\pi \frf t) dt,
    \label{eq:sp}
\end{equation}%
with $\overline{\Delta V}$ the rectified voltage, $\Delta R_{\text{STVO}}$ the device resistance variation, $I_\text{ac}$ and $\frf$ the input current amplitude and frequency, $T_1$ and $T_2$ two moments in time and $\Delta m_y$ the variation of the normalized magnetization computed from the vortex core position.
The experimental results and the DD-TEA model predictions are shown in Fig.~\ref{fig:SP_exp_vs_model}.%
%
%
\section{Results}
\begin{figure}
    \centering%
	\includegraphics[width=2.7in]{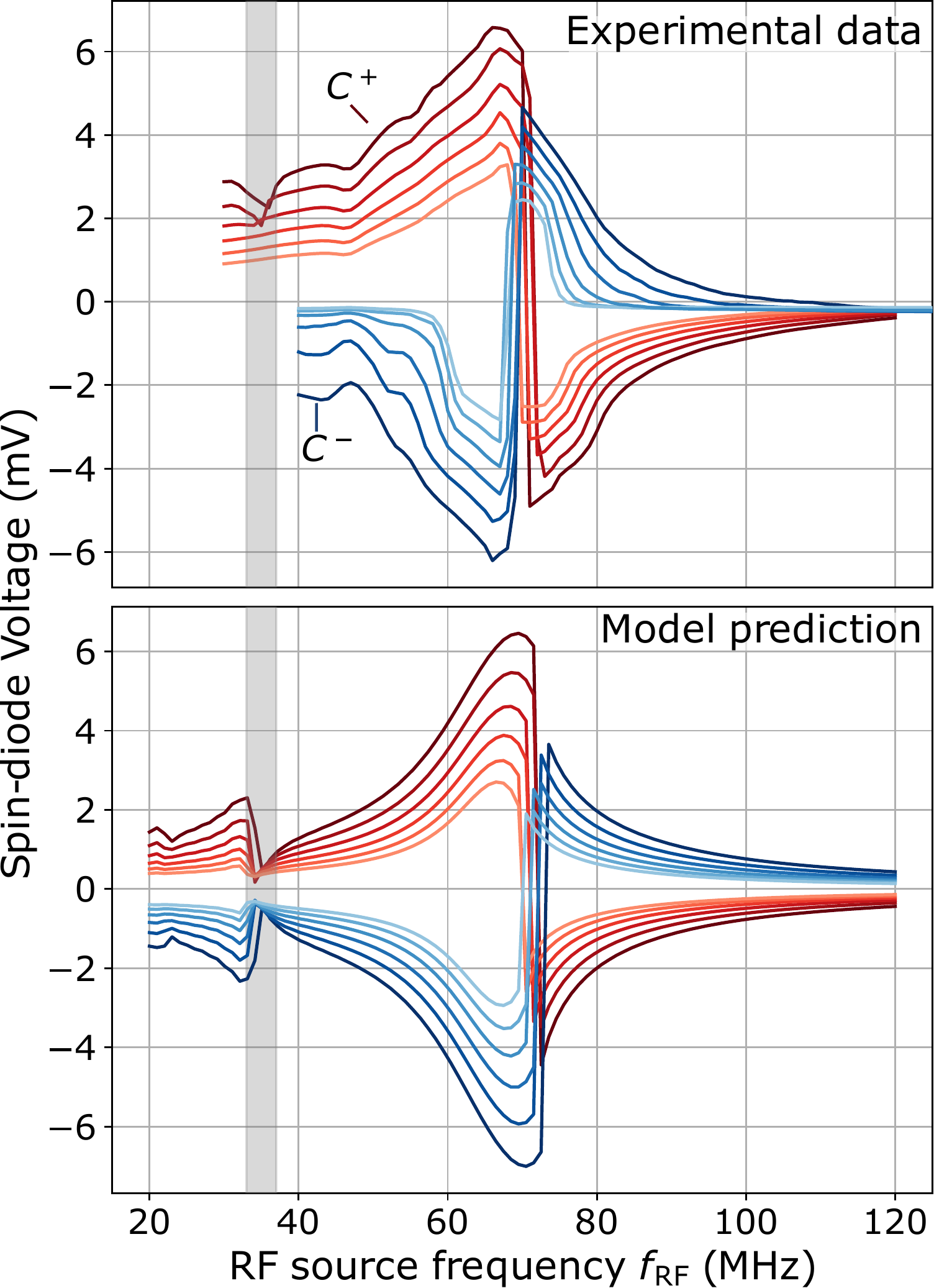}
	\caption{Spin-diode voltage as a function of the input current frequency from (top) experimental results and (bottom) model predictions. Curves in red (resp. blue) correspond to positive (resp. negative) chirality. The input power ranges between 0 dBm (darkest curve) and -5 dBm (lightest curve) with a step of -1~dBm.}
	\label{fig:SP_exp_vs_model}
\end{figure}

%
Figure~\ref{fig:SP_exp_vs_model} shows a very nice agreement between experimental data and DD-TEA predictions. %
The experimental measurements show a reversal of the spin-diode effect depending on the vortex chirality. %
This reversal comes from the impact of the AOF\cite{martins2021non} and our ultra-fast DD-TEA model is able to capture this fine-grained feature.
Also, a bump can be seen around half of the STVO frequency in the experimental data and the origin of this variation is revealed by the model. %
Thanks to its speed, the rf input frequency interval is small enough so the model offers a  high resolution. %
This allows to capture this phenomenon with, in addition, a complete knowledge of the vortex core dynamics leading to this behavior. 
It can then be shown that this behavior is due to a fractional synchronization pattern\cite{urazhdin2010fractional}. %
%
%
The combination of a STVO with the spin-diode effect can be used either to build synapses with non-volatile memory\cite{martins2021non}, as the information is stored in the vortex chirality, or as neuron with its non-linear dynamics.
Furthermore, DD-TEA can simulate a neuron with two different activation functions depending on the vortex chirality.
%
%
\section{Conclusion}
Our DD-TEA model is used to predict the vortex core dynamics.
It successfully shows and explains the spin-diode reversal effect, the impact of the input current and even explains experimental features, namely the fractional synchronization phenomenon.
These features can be used for neuromorphic computing either as neurons or synapses.
As this model is both ultra-fast and quantitative, the next step is to apply it for the simulation of full neuromorphic circuits.

\section*{Acknowledgement}
Computational resources have been provided by the Consortium des Equipements de Calcul Intensif (CECI), funded by the Fonds de la Recherche Scientique de Belgique (F.R.S.-FNRS) under Grant No. 2.5020.11 and by the Walloon Region. F.A.A. is a Research Associate and S.d.W. is a FRIA grantee, both of the F.R.S.-FNRS\@.


\end{document}